\documentstyle[12pt]{article}
\newcommand{\bc}{\begin{center}}
\newcommand{\ec}{\end{center}}
\begin{document}

{}\hfill {hep-th/9812124}

\bigskip
\bigskip

\bc {\Large\bf Effective Lagrange function \\
 of intense electromagnetic field in QED}         \ec

\bigskip
\bigskip

\bc              {\large   V.I.Ritus} \ec

\bigskip

\bc {\it Lebedev Physical Institute, 117924 Moscow, Russia} \ec

\bigskip
\bigskip

\bc              {\normalsize   Abstract} \ec

   The short survey of computation and properties of effective Lagrange
function of intensive field in two-loop approximation accounting for radiative
interaction of virtual electrons is given. The renormalization of field, charge
and mass is completely defined by the weak field behaviour of the exact
Lagrange function: its real part must be Maxwellian and imaginary part must be
quasiclassical $\propto \exp (-\pi m^2c^3/\hbar e\epsilon)$. For weak electric
field radiative interaction manifests itself as electron mass shift. Using the
renorminvariance at strong field and exponentiation at weak field it is
possible to obtain information about contributions of high order in $\alpha$.

\bigskip
\bigskip
\bigskip

\begin{center}
Published in Proceedings of the conference
             `Frontier Tests of QED and Physics of the Vacuum'
             Sandanski, Bulgaria, 9-15 June, 1998
             (Eds. E.Zavattini, D.Bakalov, C.Rizzo)
             Heron Press, Sofia, 1998
\end{center}

\newpage

  \section{Introduction}

   In 1936 Heisenberg and Euler [1] found a quantum correction to the Lagrange
function of a constant electromagnetic field. The correction takes into account
the polarization of the vacuum, i.e., the change induced by the external field
in the motion of the vacuum electrons. With this correction included, the
Lagrange function is equal to
\begin{equation}
{\cal L}=\frac{\epsilon^2-\eta^2}{2}+\frac{1}{8\pi^2}\int\limits_0^\infty
\frac{ds}{s}e^{-im^2s}\left(\frac{e^2\eta\epsilon}
 {\tan e\eta s\tanh e\epsilon s}
-\frac{1}{s^2}+\frac{e^2(\eta^2-\epsilon^2)}{3}\right).
\end{equation}
Here $\epsilon$ and $\eta$ are the intensities of the electric and magnetic
fields in the reference frame in which they are parallel. Other derivations
of the polarization correction to $\cal L$ have been proposed by Weisskopf [2]
and Schwinger [3]. The effective Lagrangian (1) is the integrand of action
$W$ and defines the vacuum conservation amplitude
\begin{equation}
e^{iW},\qquad W=\int d^4x\cal L,
\end{equation}
in the presence of the field. It has served as the basis for nonlinear quantum
electrodynamics, in which the effects are sensitive to the intensity of the
electromagnetic field and are optimal for fields $F$ of the order of $F_0=
m^2c^3/e\hbar $ (in the proper frame of electrons interacting with a field).
In (1) and below the system of units in which $\hbar=c=1$ and $\alpha=e^2/
4\pi\hbar c=1/137$ is used.

   Polarization of the vacuum is induced not only by the external
electromagnetic field but also by the individual quanta of this fields--the
photons. The photon polarization function $\pi_R $, which describes this
phenomenon and defines the exact propagator
\begin{equation}
D_R(k)=\frac{1}{k^2(1+\pi_R(k))}
\end{equation}
of a photon in vacuum, has been found (in different representations) by Serber
[4], Uehling [5], Schwinger [6,3] and Feynman [7].  It constitutes a basis for
investigations of quantum electrodynamics at small distances [8-14]. The
radiative corrections of fourth order [15-17] and (partially) sixth order [18]
have been found for the function $\pi_R $.

   It is interesting to compare the vacuum polarization by an intense field
with the vacuum polarization by quanta of large momenta, thereby linking the
intense field electrodynamics with the electrodynamics at small distances. For
this it is natural to find the next radiative corrections to Lagrange function
of an intense field and, above all, the correction associated with the change
induced by the external field in the radiative interaction of the vacuum
electrons. This correction was found by author in papers [19] and [20] for
spinor and scalar quantum electrodynamics. Its simplicity is explained by the
compactness of the Green function $G_0$ that takes the interaction with the
external field into account exactly.

\section{Lagrange function in two-loop approximation and its regularization}

   The starting equation for evaluation of Lagrange function corrections is
the equation
\begin{equation}
\delta W =\int d^4 x\: \delta A_{\alpha} \langle j_{\alpha} (x) \rangle,
\end{equation}
where the average value of current operator is defined by exact Green function
of electron
\begin{equation}
\langle j_{\alpha} (x)\rangle=-ie\: \hbox{tr}\: \gamma_{\alpha} (x\vert G\vert
x), \end{equation} while the change of external field potential is connected
with Green function $G_0$ without radiation corrections:  \begin{equation}
e\gamma\delta A=\delta G_0^{-1}.
\end{equation}

   Then after the integration of $\delta W$ with these ingredients it appears
that the first-order nonlinear correction to Maxwell Lagrange function is
equal to
\begin{equation}
{\cal L}^{(1)}(x)=i\:\hbox{tr}\:(x\vert\ln G_0\vert x)
\end{equation}
and corresponds to one-loop diagram 1 in the figure, while the second-order
nonlinear correction is equal to
\begin{equation}
{\cal L}^{(2)}(x)=\frac 12\:\hbox{tr}\:(x\vert M^{(2)}G_0\vert x)
\end{equation}
and is characterized by two two-loop diagrams 2a and 2b in the figure.
The double line in this figure represents a virtual electron interacting
with the external field, and the wavy line represents a virtual photon. In the
case of a constant and uniform external field, which is considered here, the
tadpole diagram 2b makes no contribution to ${\cal L}^{(2)}$ since the average
current induced in the vacuum by such a field is equal to zero. Thus, in this
case the polarization correction ${\cal L}^{(2)}$ takes into account the
interaction of the electron and positron of one virtual pair.

   To calculate
\begin{equation}
{\cal L}^{(2)}=\frac 12 ie^2\int d^4x'\:\hbox{tr}\:[\gamma_{\mu}G_0(x,x')
\gamma_{\mu}G_0(x',x)]\:D_0(x-x')
\end{equation}
we shall make use of the Green function of an electron in a constant field,
found by Fock [21] and Schwinger [3]:
\begin{eqnarray}
G_0(x,x')=\frac {-i\exp [i\phi(x,x')]}{16\pi^2}\int\limits_0^\infty
\frac{ds}{s^2}\left [m-\frac i2\gamma(\beta+eF)z\right ]\nonumber\\
\times \exp\left[-im^2s-L(s)+\frac{iz\beta z}{4}+\frac{ie\sigma Fs}{2}\right ].
\end{eqnarray}
Here
$$
\beta=eF\:\coth eFs,\qquad L(s)=\frac 12\:\hbox{tr}\:\ln [(eFs)^{-1}\sinh eFs]
$$
are the matrix and scalar functions of the matrix $F,\;z=x-x',$ and $\phi $ is
the off-diagonal phase of the Green function, equal to the line integral of
potential $eA_{\mu}(y)$ along the straight line joining the points $x'$ and
$x$. Finally,
\begin{equation}
D_0(z)=\frac {-i}{16\pi^2}\int\limits_0^\infty \frac{dt}{t^2}\exp \left (
\frac{iz^2}{4t}\right )
\end{equation}
is the photon Green function in proper-time representation.

   After integration over space-time and over photon proper-time and
regularization with the help of small electron proper-time $s_0$ the
${\cal L}^{(2)}$ is represented in the following form:
\begin{equation}
{\cal L}^{(2)}=-{\cal L}^{(0)}\frac {\alpha^2}{4\pi^2}\left (\ln\frac {1}
{i\gamma m^2 s_0}-\frac 53+\frac 53 \ln 2\right )+\delta m^2\frac{\partial%
{\cal L}_R^{(1)}}{\partial m^2}+{\cal L}_R^{(2)},
\end{equation}
where
\begin{equation}
{\cal L}_R^{(2)}=\frac {-i\alpha}{16\pi^3}\int\limits_0^\infty ds\left\{%
\int\limits_0^s ds'\left [K(s,s')-\frac{K_0(s)}{s'}\right ]+K_0(s)\:\left (%
\ln i\gamma m^2 s-\frac56\right )\right\},
\end{equation}
and
\begin{equation}
\delta m^2=\frac{3\alpha m^2}{2\pi}\left (\ln\frac{1}{i\gamma m^2 s_0}+
\frac 56\right )
\end{equation}
is the electromagnetic part of the square of the electron mass, so that
$m^2=m_0^2+\delta m^2$ is the square of the real electron mass. The functions
$K(s,s')$ and $K_0(s)$ are given in Appendix.

   At the same time, the first-order radiative correction to Lagrange function
has the form
\begin{equation}
{\cal L}^{(1)}={\cal L}^{(0)}\frac {\alpha_0}{3\pi}\ln\frac{1}{i\gamma m^2_0%
s_0}+{\cal L}_R^{(1)}(m_0^2),
\end{equation}
where ${\cal L}_R^{(1)}$ is determined by formula (1) and the unrenormalized
fine-structure constant and electron mass are indicated by the subscript zero.

\section{Charge and mass renormalization}

   Then the unrenormalized Lagrange function
$$
{\cal L}^{(0)}=\frac12 (\epsilon^2_0-\eta^2_0)
$$
of the Maxwell field and the terms proportional to it in the
expressions (15) and (12) are collected into the renormalized Lagrange function
$$
{\cal L}_R^{(0)}=\frac12 (\epsilon^2-\eta^2),
$$
namely,
\begin{eqnarray}
{\cal L}_R^{(0)}={\cal L}^{(0)}Z_3^{-1},\hspace{3cm} \nonumber\\
Z_3^{-1}=1+\frac{\alpha_0}{3\pi}\ln\frac 1{i\gamma m_0^2 s_0}-\frac{\alpha
^2_0}{4\pi^2}\left (\ln\frac 1{i\gamma m^2_0 s_0}-\frac 53+\frac 53\ln 2
\right ),
\end{eqnarray}
thereby leading to renormalization of the field and charge:
\begin{equation}
\eta =\eta_0 Z_3^{-1/2},\qquad \epsilon =\epsilon_0 Z_3^{-1/2},\qquad
e=e_0 Z_3^{1/2}
\end{equation}
(we have indicated the unrenormalized values of the field and charge by the
subscript zero), and the term $\delta m^2\partial {\cal L}_R^{(1)}/\partial
m^2$ in (12) effects the renormalization of the electron mass:
\begin{equation}
{\cal L}_R^{(1)}(m_0^2)+\delta m^2\frac {\partial {\cal L}_R^{(1)}(m_0^2)}
{\partial m_0^2}={\cal L}_R^{(1)}(m^2);
\end{equation}
the functions ${\cal L}_R^{(1)}(m^2)$ and ${\cal L}_R^{(2)}(m^2)$ are
determined by formulas (1) and (13). The arguments $e_0\eta_0=e\eta,\;
e_0\epsilon_0=e\epsilon$ of these functions are invariant under the
renormalization (17). The factor $Z_3^{-1}$ can be expressed by means of (14)
in terms of the renormalized electron mass:
\begin{equation}
Z_3^{-1}=1+\frac{\alpha_0}{3\pi}\ln \frac 1{i\gamma m^2 s_0}+\frac{\alpha_0^2}
{4\pi^2}\left (\ln\frac 1{i\gamma m^2 s_0}+\frac {10}{3}-\frac 53\ln 2\right ).
\end{equation}
The coefficients of the logarithms in this expression coincide with those
known in literature; cf., e.g.,[14].

   Thus, up to radiative corrections of second order in $\alpha $, the
nonlinear Lagrange function of a constant field is equal to
\begin{equation}
{\cal L}
={\cal L}^{(0)}+{\cal L}^{(1)}+{\cal L}^{(2)}={\cal L}_R^{(0)}+{\cal L}_R
^{(1)}(m^2)+{\cal L}_R^{(2)}(m^2),
\end{equation}
in which the latter expression contains only the renormalized values of the
electron charge and mass.

   Note, that because of the absence of mass parameter in ${\cal L}^{(0)}$
the necessity of mass renormalization appears only at consideration of two-loop
and higher corrections to ${\cal L}^{(0)}$.

\section{Asymptotic properties of ${\cal L}^{(2)}$, connection with
    QED at small distances}

   We shall find the limiting expressions for the Lagrange function ${\cal L}
_R^{(2)}$ for weak and strong fields, and consider its imaginary part in detail.

   Expanding the functions $K(s,s')$ and $K_0(s)$ in the field and integrating
these expansions over the proper times $s$ and $s'$, we obtain, according to
formula (13),
\begin{equation}
{\cal L}_R^{(2)}=\frac{\alpha ^3}{\pi m^4}\left [\frac{16}{81}(\eta^2-\epsilon
^2)^2+\frac{263}{162}(\eta\epsilon)^2\right ]+\ldots ;\quad \frac{e\eta}{m^2},
\:\frac{e\epsilon}{m^2}\ll 1.
\end{equation}
In expression (21) the imaginary part of ${\cal L}_R^{(2)}$, which is
exponentially small compared with the real part and cannot be represented by a
series in powers of the field, is absent. In general, the function ${\cal L}
_R^{(2)}$ has an essential singularity at the point $eF=0$, so that the series
(21) for the real part of ${\cal L}_R^{(2)}$ is asymptotic.

   In the case of a strong magnetic field the calculations are conveniently
carried out by first rotating the contours of integration over the proper times
: $s, s'\to -is, -is'$. We then obtain, with logarithmic accuracy,
\begin{equation}
{\cal L}_R^{(2)}\approx \frac{\alpha^2\eta^2}{8\pi^2}\left (\ln\frac{e\eta}
{\gamma \pi m^2}+a_2\right ), \qquad \frac{e\eta}{m^2}\gg 1, \;
\frac{e\epsilon}{m^2}\le 1.
\end{equation}
The uncalculated constant $a_2$, additive to logarithm, is real.

   The case of a strong electric field can be obtained from (22) by the
transformation $\eta \to -i\epsilon$ :
\begin{equation}
{\cal L}_R^{(2)}\approx -\frac{\alpha^2\epsilon^2}{8\pi^2}\left (\ln
\frac{e\epsilon}{\gamma \pi m^2}-\frac{i\pi}{2}+a_2 \right), \qquad
\frac{e\epsilon}{m^2}\gg 1, \; \frac{e\eta}{m^2}\le 1.
\end{equation}
In this way the correct asymptotic expression for the imaginary part of
${\cal L}_R^{(2)}$ is obtained.

   It is useful to compare formulas (21) and (22) with the corresponding
formulas for ${\cal L}_R^{(1)}$ :
\begin{eqnarray}
{\cal L}_R^{(1)}\approx \frac{2\alpha^2}{45m^4}[(\eta^2-\epsilon^2)^2+7(\eta
\epsilon )^2 ], \qquad \frac{e\eta}{m^2}, \; \frac{e\epsilon}{m^2}\ll 1,
\hspace{5mm} \nonumber\\
{\cal L}_R^{(1)}\approx \frac{\alpha\eta^2}{6\pi}\left(\ln\frac{e\eta}
{\gamma \pi m^2}+\frac{6}{\pi^2}\zeta'(2)\right), \qquad \frac{e\eta}
{m^2}\gg 1, \; \frac{e\epsilon}{m^2}\le 1.
\end{eqnarray}
Here $6\pi^{-2}\zeta'(2)=-0.5699610\ldots ; \zeta (x)$ is the Riemann
zeta-function. As can be seen from the asymptotic formulas obtained, the
function ${\cal L}_R^{(2)}\sim \alpha {\cal L}_R^{(1)}$, irrespective of the
magnitude of the field. In Sec.8 it will be shown that this is an exception,
and the ratio of the subsequent radiative corrections will be of order
$(\alpha /\pi )\ln(eF/m^2)$ for large fields.

   A remarkable property of the functions ${\cal L}_R^{(1)}$ and ${\cal L}_R
^{(2)}$ is the fact that for large magnetic (electric) fields the ratios
\begin{eqnarray}
\frac{{\cal L}_R^{(1)}}{{\cal L}_R^{(0)}}\approx \frac{\alpha}{\pi}\left (
-\frac13\ln\frac{e\eta}{\gamma\pi m^2}-\frac{2}{\pi^2}\zeta'(2)\right ),
\nonumber\\
\frac{{\cal L}_R^{(2)}}{{\cal L}_R^{(0)}}\approx \left(\frac{\alpha}{\pi}
\right)^2\left(-\frac14\ln\frac{e\eta}{\gamma\pi m^2}+a_{20}\right),
\end{eqnarray}
behave, with logarithmic accuracy, in the same way as the polarization
functions of second and fourth order in $e$ at large values of the squares
of the space-like (time-like) momenta
\begin{eqnarray}
\pi_R^{(2)}\approx \frac{\alpha}{\pi}\left(-\frac13\ln\frac{k^2}{m^2}+\frac59
\right),\hspace{1cm} \nonumber\\
\pi_R^{(4)}\approx \left(\frac{\alpha}{\pi}\right)^2\left(-\frac14\ln\frac{k^2}
{m^2}+\frac{5}{24}-\zeta(3)\right).
\end{eqnarray}

   By the nature of the problem, the Lagrange function and the polarization
function describe the same phenomenon and are determined by the effective value
of the operator $\Pi^2, \Pi_\alpha =p_\alpha-eA_\alpha $, responsible for the
interaction of the vacuum electrons with the quanta or with the field. At large
momenta of the quanta, or high fields, the average value of $\Pi^2$ becomes of
the order of $k^2\gg m^2$ for the quanta and of the order of $(eFx)^2\sim eF
\gg m^2$ for the field, since in this case the Lagrange function is formed
over distances $x\sim (eF)^{-1/2}$ that are short compared with the Compton
wavelength.

   Therefore, the Lagrange function of the constant field also correctly
describes the polarization corrections for varying fields, if the fields are
sufficiently intense. For example, the well-known Uehling-Serber correction to
the Coulomb field at short distances:
\begin{equation}
\epsilon=\frac{q}{4\pi r^2}\left[1+\frac{\alpha}{3\pi}\left(\ln\frac{1}
{(\gamma mr)^2}-\frac53\right)\right], \qquad r\ll m^{-1},
\end{equation}
calculated with the aid of the polarization function $\pi_R$, can also be
found, with logarithmic accuracy, with the aid of the Lagrange function
$$
{\cal L}=\frac{\epsilon^2}{2}-\frac{\alpha\epsilon^2}{6\pi}\ln\frac{e\epsilon}
{\gamma\pi m^2}
$$
of a constant electric field, if we solve for $\epsilon$ the expression for the
induction
\begin{equation}
\frac{q}{4\pi r^2}=\frac{\partial{\cal L}}{\partial\epsilon}\approx \epsilon
\left(1-\frac{\alpha}{3\pi}\ln\frac{e\epsilon}{\gamma\pi m^2}\right),
\end{equation}
which is equal to $q/4\pi r^2$ since its divergence is equal to zero.

\section{The virial representation of the imaginary part of the
    Lagrange function}

   We consider now the imaginary part of the Lagrange function ${\cal
L}^{(2)}$. We recall, in this connection, that $e^{i{\cal L}VT}$ is the
amplitude and $\vert e^{i{\cal L}VT}\vert^2=\exp(-2\hbox{Im}{\cal L}VT)$ is the
probability of vacuum conservation in volume $V$ for a time $T$.

   For the imaginary part of the Lagrange function of a constant and
homogeneous electric field $\epsilon$ in the one-loop approximation Schwinger
[3] obtained the representation
\begin{equation}
2\hbox{Im}{\cal L}^{(1)}=(2s+1)\frac{(e\epsilon )^2}{(2\pi)^3}\sum_{n=1}^\infty
\frac{(\pm 1)^{n+1}}{n^2}e^{-\pi n/\beta}, \qquad \beta=\frac{e\epsilon}{m^2}.
\end{equation}
Here and below, the upper and lower signs correspond to charged Fermi and Bose
particles with spin $s=1/2$ and $s=0$. As can be seen from this representation,
the decay rates of the vacua of Fermi and Bose fields differ under otherwise
equal conditions by not only the spin statistical weight (2s+1). To elucidate
the physical meaning of the individual terms of the representation (29), it is
convenient to use Nikishov's representation [22]
\begin{eqnarray}
\frac{2}{\hbar}\hbox{Im}{\cal L}^{(1)}VT=\mp \sum_r\int\frac{d^3pV}
{(2\pi\hbar)^3}\ln(1\mp \bar n_p), \nonumber\\
\bar n_p=\exp\left(-\pi\frac{m^2+p^2_\bot}{e\epsilon}\right), \hspace{15mm}
\end{eqnarray}
which relates the imaginary part of the Lagrange function of the field to the
mean number $\bar n_p$ of pairs produced by the field in the state with given
momentum and spin projection $p=\vec p, r$. The distribution $\bar n_p$ is
degenerated with respect to the spin projection $r$ and the momentum component
$p_{\Vert}$ longitudinal with respect to the field with degeneracies $2s+1$
and $L_{\Vert}\Delta p_{\Vert}/2\pi\hbar$, where $\Delta p_{\Vert}=e\epsilon
T$ (see [22,23]).

   It was pointed out in [24] that the right-hand side of the representation
(30) as a functional of $\bar n_p$ is equal to $PV/\Theta $ for an ideal Fermi
or Bose gas, where $P, V,$ and $\Theta$ are the pressure, volume, and
temperature. Then $2\hbox{Im}{\cal L}^{(1)}/\hbar $ is analogous to the rate of
growth of the pressure in units of $\Theta $.

   Expansion of $\ln(1\mp\bar n_p)$ in (30) in a series in powers of $
\bar n_p$ and term-by-term integration over $p_\bot $ leads to the series (29).
Therefore, the first term of the series (29) is the mean number $\bar n $ of
pairs in unit 4-volume [22], and the approximate expression
\begin{equation}
2\hbox{Im}{\cal L}^{(1)}\approx \bar n=(2s+1)\frac{(e\epsilon)^2}{(2\pi)^3}
e^{-\pi/\beta}
\end{equation}
is analogous to the equation of state of an ideal gas (Clapeyron's equation).
The following terms of the series (29) with $n\ge 2$ are the quantum-mechanical
exchange corrections to the rate of growth of the pressure describing the
additional Fermi-repulsion or Bose-attraction of the particles for the given
mean 4-density of their number. They arise because of coherent pair creation,
i.e., creation of $n\ge 2$ pairs in the same 4-volume of pair formation.

   Therefore, (29) is a virial expansion for $2\hbox{Im}{\cal L}^{(1)}$ in
powers of the parameter
$$
\exp (-\pi/\beta)\approx \bar n(e\epsilon)^{-2},
$$
which is the mean number of pairs in the 4-volume of pair formation [24]. Such
a representation is analogous to the virial expansion of the pressure of an
ideal Fermi or Bose gas in powers of the degeneracy parameter, i.e., the mean
number of particles in the 3-volume determined by the thermal de Broglie
wavelength (see \S 56 in [25]).

   In statistical physics the virial expansion of the pressure proved to be
very convenient for describing the deviation of the state of a gas from the
ideal state not only due to the effects of quantum-mechanical degeneracy but
also the force interaction of molecules. In our case, the force, i.e.,
radiative, interaction of the vacuum charges is taken into account by the
two-loop correction ${\cal L}^{(2)}$ to the Lagrange function of the
electromagnetic field, which was found in [19,20] in proper-time
representation.

   In the paper by Lebedev and author [26] the imaginary part of the Lagrange
function of a constant electric field in the two-loop approximation was
represented in the form (29), i.e., for $2\hbox{Im}\cal L$ we obtain with
allowance for the radiative interaction of the vacuum charges the virial
representation
\begin{eqnarray}
2\hbox{Im}({\cal L}^{(1)}+{\cal L}^{(2)})= \hspace{3cm} \nonumber\\
=(2s+1)\frac{(e\epsilon)^2}{(2\pi)^3}\sum_{n=1}^\infty(\pm 1)^{n+1}\left[
\frac{1}{n^2}+\alpha\pi K_n(\beta, s)\right]e^{-n\pi/\beta}.
\end{eqnarray}
The complicated function $K_n(\beta,s)$ simplifies for $n\beta\ll 1$ and in
this region does not depend on the spin:
\begin{eqnarray}
K_n(\beta,s)=-\frac{c_n}{\sqrt\beta}+1+\ldots, \hspace{3cm} \nonumber\\
c_1=0,\qquad c_n=\frac{1}{2\sqrt n}\sum_{k=1}^{n-1}\frac{1}{\sqrt{k(n-k)}},
\quad n\ge2,
\end{eqnarray}
the dots denote terms that vanish as $\beta\to 0$. For $n=1$, this result was
found earlier by author in [19] and [20].

\section{ Renormalization principle immanent to Lagrange function}

   Below I give a physical interpretation of the radiation correction (33) on
the basis of the independent physical meaning of the terms in the virial
expansion, which gives a picture of coherent pair production. This meaning of
the individual terms of the virial series makes it possible to transform the
radiative correction into a field-dependent correction to the mass of a charge
participating in group tunneling. Such a transformation is equivalent to
summation of the radiative corrections of all orders in $\alpha$ in a weak
field. Moreover, in all terms of the virial expansion the mass of the charge
does not acquire finite corrections (non-vanishing with the field), this being
due to the correct renormalization of the mass of the charge and to the
quasiclassical meaning of the $n$th term of the virial expansion for a weak
field (with regard to the quasiclassical nature of terms of the series (29),
see [27]).

   Indeed, it was shown in [19] and [20] that by means of the Lagrange function
one can uniquely renormalize the charge and mass of the particles by requiring
that in the weak-field limit the real part of the exact Lagrange function be
Maxwellian and the imaginery part be quasiclassical:
\begin{equation}
\hbox{Im}{\cal L}=(e\epsilon)^2f(\alpha)e^{-\pi m^2/e\epsilon},\quad \epsilon
\to 0.
\end{equation}
Here, $m$ is the renormalized observable mass of the charged particles. In
accordance with this condition, the radiative corrections in the weak field
change only the pre-exponential factor $f(\alpha)=f^{(1)}+\alpha f^{(2)}+
\alpha^2f^{(3)}+\ldots$, and do not change the argument of the exponential,
which by virtue of the quasiclassical situation plays in the mass
renormalization a part similar to the part played by the classical Thomson
cross section in charge renormalization, namely, in the limit $\epsilon\to 0$
the radiative corrections to $m^2$ tend to zero like the radiative corrections
to the cross section of the Compton effect when the photon frequency tends to
zero, $\omega\to 0$; see [28].

   Thus, the boundary condition (34) imposed on the imaginary part of the exact
Lagrange function uniquely fixes the mass renormalization [19]. Because the
exponentials with $n\ge 2$ are quasiclassical, the singularities of the type
$C\beta^{-1},\;n\beta\ll 1$, disappear after the mass renormalization in all
terms of the virial expansion with $n\ge 2$. Such singularities would lead
after exponentiation to a finite, nonvanishing as $\beta\to 0$, mass shift in
the terms with $n\ge 2$, and this would be in contradiction with a
quasiclassical nature of the exponentials, where $m$ is defined as the physical
mass of the particle [19]. Thus, to fix the mass renormalization uniquely, a
boundary condition of the type (34) could be imposed on any term of the virial
series. This emphasizes once more the independent physical meaning of each of
these terms.

\section{ Interpretation of the field dependent radiative corrections to
electron mass}

   The radiative correction to the $n$-th term of the virial expansion for
$n\beta\ll 1$ has the following remarkable properties, see (33):

   1) it does not depend on the spin $s$;

   2) for $n\ge 2$ it has a term singular in the limit $\beta\to 0$, in
contrast to the correction to the leading term $n=1$;

   3) the singular and the constant terms have different signs;

   4) for $n\gg 1$ the correction depends on $n$ and $\beta$ only through the
product $n\beta$, since $c_n\approx (\pi/2)n^{-1/2}$ for $n\gg 1$.

   All these properties obtain a simple physical explanation.

   Since the terms of the virial expansion have an independent physical
meaning, the radiative corrections to them must not increase unboundedly with
decreasing field. This suggests that allowance for the further corrections of
order $\alpha^2,\,\alpha^3,\ldots$, will lead to exponentiation of the
correction in (33):
\begin{eqnarray}
\left[\frac{1}{n^2}+\alpha\pi\left(-\frac{c_n}{\sqrt\beta}+1+\ldots\right)
\right]\exp\left(-\frac{n\pi m^2}{e\epsilon}\right)
\approx\frac{1}{n^2}\exp\left( -\frac{n\pi m^2_*}{e\epsilon}\right),
\end{eqnarray}
where
\begin{equation}
 m_*(n)=m+\frac 12\alpha nc_n\sqrt{e\epsilon}-\frac 12\alpha n\frac{e\epsilon} {m},\qquad
n\beta\ll 1,
\end{equation}
i.e., $2\hbox{Im}({\cal L}^{(1)}+{\cal L}^{(2)}+\ldots )$ acquires the form of
the virial series (29) with replacement of the mass $m$ in the weak field by the
mass $m_*(n)$ determined by the expression (36).

   The properties of the radiative correction listed above are now transformed
into properties of the mass shift (36) and can be explained by means of a
picture of the coherent production of $n$ pairs [24]. The coherent tunneling
of $n$ particles separated from each other by the longitudinal distance $\Delta
 r_{\Vert}\sim m^{-1}$ takes place as the tunneling of one particle with charge
$Q=ne$ and mass $M=nm$ through a barrier of width $2M/Q\epsilon=2m/e\epsilon$.
Therefore, to coherent tunneling of $n$ particles there corresponds the
quasiclassical exponential $\exp(-n\pi/\beta)$, and their distribution with
respect to the transverse momenta has the form $(\bar n_p)^n$. Consequently,
the effective value of the transverse momentum of a particle in the group is
$p_\bot\sim\sqrt{e\epsilon/n}$, and the effective transverse distance between
the particles is $r_\bot\sim\sqrt{n/e\epsilon}$. As a result of the coherent
tunneling two narrow groups of oppositely charged particles are formed during
the time $\Delta t\sim m/e\epsilon$ with distance $r_\Vert\approx 2m/e\epsilon$
between them:
\begin{eqnarray}
\Delta r_{\Vert}\sim\frac nm\ll r_{\bot}\sim\sqrt{\frac{n}{e\epsilon}}\ll
\frac{2m}{e\epsilon}.
\end{eqnarray}
The condition (37) is equivalent to the condition $n\beta\ll 1$, under which
the asymptotic behavior (33) is valid.

   The negative term $-(1/2)\alpha n(e\epsilon/m)$ of the mass shift (36) does
not depend on $\hbar$ and can be interpreted as the energy of effective
attraction of an individual charge to the $n$ charges of the opposite sign
in the coherent group at distance $r_\Vert\approx 2m/e\epsilon$ from it:
\begin{equation}
\left.-\frac{ne^2}{4\pi r_\Vert}\right\vert_{r_\Vert=2m/e\epsilon}=-\frac12
\alpha n\frac{e\epsilon}{m}.
\end{equation}
For $n=1$, the negative term of the shift (36) exhausts the entire shift and
is equal to the classical mass shift of an accelerated charge found by author
in [29].

   The positive term $(1/2)\alpha nc_n\sqrt{e\epsilon}$ of the shift (36)
depends on $\hbar$ and is the energy of the Coulomb repulsion of an individual
charge by the remaining $n-1$ like charges, the partners in the coherent group,
which are at distance $r_\bot\sim\sqrt{n\hbar c/e\epsilon}$ from it (see (37)):
\begin{equation}
\left.\frac{(n-1)e^2}{4\pi r_\bot}\right\vert_{r_\bot \sim\sqrt{n\hbar c/e
\epsilon}}\sim \alpha\frac{n-1}{\sqrt n}\sqrt{\hbar ce\epsilon}.
\end{equation}
The estimates (38) and (39) reproduce all the qualitative features of the
shift (36), including the fact that it does not depend on the spin.

   Particular attention should be drawn to the agreement between the mass shift
for the leading term $n=1$ and the mass shift of a uniformly accelerated charge
[29]. This casts light on the reason for the radiative enhancement of pair
production by a week electric field: the accelerated charges are lighter than
the unaccelerated charges, and therefore they can be more readily produced.
In addition, this also makes it possible to give a new interpretation of the
shift itself. Whereas in the scattering channel ``$e$+field $\to e$+field" it
can be interpreted as a reactive energy [30], or as a manifestation of the clock
paradox [31], or as a red shift of the proper field energy of accelerated
charge [32],
in the cross channel ``field $\to e^+e^-+$field" it can be regarded as an
effective attraction between the $e^+e^-$ in their region of formation.

\section{Lagrange function of an intense field from the viewpoint of
    the renormalization group}

   The relation (20) for the Lagrange function expresses its invariance with
respect to renormalization of the field, charge and mass:
\begin{equation}
{\cal L}(e_0F_0, \alpha_0, m^2_0, im^2_0s_0)={\cal L}_R(eF, \alpha, m^2).
\end{equation}
If in the left-hand side of (40) we carry out the mass renormalization and
use the invariance of the product $e_0F_0=eF$, this relation takes the form
\begin{equation}
{\cal L}(eF, \alpha_0, m^2, im^2s_0)={\cal L}_R(eF, \alpha, m^2),
\end{equation}
with
$$
\alpha=\alpha_0 Z_3(im^2s_0, \alpha_0),
$$
cf. (17) and (19). By dividing both sides of the equality (41) by the
renorm-invariant function
$$
\alpha{\cal L}^{(0)}_R\equiv -\frac{(eF)^2}{16\pi}
$$
and denoting the ratios ${\cal L}/{\cal L}^{(0)}$ and
${\cal L}_R/{\cal L}^{(0)}_R$ by $l$ and $l_R$, we obtain in place of (41)
an invariance relation for the dimensionless functions:
\begin{equation}
\alpha_0^{-1}l\left(\frac{eF}{m^2}, \alpha_0, im^2s_0\right)=\alpha^{-1}
l_R\left(\frac{eF}{m^2}, \alpha \right).
\end{equation}
The function $\alpha^{-1}l_R$ may be called the inverse square of the
renorm-invariant charge.

   We shall assume that for $eF/m^2\to\infty$ the function $l$ has an
asymptotic form that does not depend on $m^2$, namely the $\lim l=l_\infty
(ieFs_0, \alpha_0)$. More precisely, we assume that when one of the field
parameters tends to infinity, e.g., $e\eta/m^2\to\infty$, the asymptotic
form of $l$ depends neither on $m^2$ nor on the second parameter $e\epsilon$.
This masslessness assumption is fulfilled for the function $l=1+l^{(1)}+
l^{(2)}$ in the approximation found (cf. (12) and (15)). Then, substitution
into $l_\infty $ the quantity $im^2 s_0=\varphi (\alpha, \alpha_0 )$ found
from the relation $\alpha=\alpha_0 Z_3(im^2s_0, \alpha_0 )$, we obtain,
according to (42),
\begin{equation}
\alpha_0^{-1} l_\infty\left(\frac{e\eta}{m^2}\varphi (\alpha, \alpha_0 ),
\alpha_0\right)=\alpha^{-1} l_{R\infty}\left(\frac{e\eta}{m^2},\alpha\right).
\end{equation}

   Inasmuch as the right-hand side does not depend on $\alpha_0$, the left-hand
side should also not depend on $\alpha_0$. Consequently, the left-hand side has
the form $\Phi [e\eta m^{-2}\varphi (\alpha )]$, i.e., is a function of one
variable. This means that the function
\begin{equation}
\alpha^{-1} l_{R\infty}\left(\frac{e\eta}{m^2}, \alpha \right)=\Phi \left[
\frac{e\eta}{m^2}\varphi (\alpha)\right]
\end{equation}
satisfies the Callan-Symanzik equation [33,34]
\begin{eqnarray}
\left(m^2\frac{\partial}{\partial m^2}+
\alpha\beta(\alpha)\frac{\partial}{\partial
\alpha}\right)\alpha^{-1}l_{R\infty}\left(\frac{e\eta}{m^2},\alpha\right)=0,
\nonumber\\
\beta(\alpha)=\frac{\varphi (\alpha)}{\alpha\varphi'(\alpha)}=
\frac{m^2dZ_3}{Z_3dm^2}. \hspace{2cm}
\end{eqnarray}
The derivative $dZ_3/dm^2$ is calculated at fixed $s_0$ and $\alpha_0$.

   Note that differentiating equality (42) with respect to $m^2$ with fixed
values of $eF, \alpha_0$ and $s_0$ we may obtain the Callan-Symanzik equation
for nonasymptotic function $\alpha^{-1}l_R$:
\begin{equation}
\left(m^2\frac{\partial}{\partial m^2}+\alpha\beta(\alpha)\frac{\partial}
{\partial\alpha}\right)\alpha^{-1}l_R\left(\frac{eF}{m^2},\alpha\right)=
m^2\frac{\partial}{\partial
m^2}\left[\alpha_0^{-1}l\left(\frac{eF}{m^2},\alpha_0,im^2s_0\right)\right].
\end{equation}

If in the asymptotic domain $eFm^{-2}\gg 1$ the masslessness assumption is
valid for $l$, then in this region the right half of (46) vanishes and the
equation goes over to (45).

   According to perturbation theory, the function $l_{R\infty}$ can be
expanded in the following series in powers of $\alpha$:
\begin{equation}
l_{R\infty}=1+\frac{\alpha}{\pi}(a_{10}+a_{11}z)+\sum_{n=2}^\infty \left(
\frac{\alpha}{\pi}\right)^n \sum_{k=0}^{n-1}a_{nk}z^k,\quad z=\ln\frac{e\eta}
{\gamma\pi m^2}.
\end{equation}
Its coefficients $a_{nk}$ are, in accordance with (45), related to each other
and to the coefficients of the power series for $\beta (\alpha)$
\begin{equation}
\beta (\alpha)=\sum_{k=1}^\infty \beta_k\left(\frac{\alpha}{\pi}\right)^k=
\frac13\left(\frac{\alpha}{\pi}\right)+\frac14\left(\frac{\alpha}{\pi}\right)
^2-\frac{121}{288}\left(\frac{\alpha}{\pi}\right)^3+\ldots
\end{equation}
by the relations
\begin{equation}
a_{00}=1,\quad a_{n1}=\sum_{i=0}^{n-1}(i-1)a_{i0}\beta_{n-i},\quad
ka_{nk}=\sum_{i=k}^{n-1}(i-1)a_{ik-1}\beta_{n-i},\quad k\ge 2.
\end{equation}

   It follows from these relations that in terms of order $\alpha^n,\; n\ge2$,
the exponent of the highest power of the logarithm will be $n-1$. The
coefficients of the highest powers of the logarithms in (47) are determined
entirely by the coefficients $\beta_1$ and $\beta_2$:
\begin{equation}
a_{11}=-\beta_1,\quad a_{nn-1}=-\frac{\beta_2\beta_1^{n-2}}{n-1},
\end{equation}
while the coefficients $a_{nn-r},\; 2\le r\le n-1$, of the other logarithms are
determined by the coefficients $\beta_1,\:\beta_2,\ldots ,\:\beta_{r+1}$ and
the constants $a_{20},\: a_{30},\dots ,\: a_{r0}$. The latter constants,
together with $a_{10}$, reflect the structure of the function
$\Phi$ and its arbitrariness--the arbitrariness of the renormalization group:
the function $l_{R\infty}$ is characterized by fully determined values of its
constants $a_{n0}$ if $e=\sqrt{4\pi\alpha}$ and $m$ are the charge and mass of
a real electron. Other values of $a_{n0}$ will correspond to a different
meaning of the parameters $e$ and $m$. On the other hand, only by the constants
$a_{n0}$ can the asymptotic form $l_{R\infty}$ of the Lagrange function at
large fields be distinguished from the asymptotic form $\pi_{R\infty}$ of the
polarization function at large momenta.

   Using (50) and formulas not written out here for the coefficients
$a_{nn-r}$ of the logarithms of lower rank $r,\: 2\le r\le n-2$, we can find
the contribution to the sum (47) from the logarithms of the first, second,
etc., ranks $r=1,2,\ldots$ in all orders in $\alpha$:
\begin{eqnarray}
\left(\frac{\alpha}{\pi}\right)^rL_r=\sum_{n=r+1}^\infty \left(\frac{\alpha}
{\pi}\right)^n a_{n,n-r}z^{n-r},\hspace{2cm} \nonumber\\
L_1=\frac{\beta_2}{\beta_1}\ln(1-x),\qquad x=\frac{\alpha}{\pi}\beta_1 z,
\hspace{5cm} \nonumber\\
L_2=\left(\frac{\beta_2}{\beta_1}\right)^2\frac{\ln(1-x)+x}{1-x}+\left(a_{20}-
\frac{\beta_3}{\beta_1}\right)\frac{x}{1-x},\hspace{2,5cm} \nonumber\\
L_3=-\left(\frac{\beta_2}{\beta_1}\right)^3\frac{\ln^2(1-x)-x^2}{2(1-x)^2}+
\frac{\beta_2\beta_3}{\beta_1^2}\frac{\ln(1-x)+x(1-x)}{(1-x)^2}-\nonumber\\
-\frac{\beta_2a_{20}}{\beta_1}\frac{\ln(1-x)}{(1-x)^2}+\left(a_{30}-
\frac{\beta_4}{2\beta_1}\right)\frac{x(2-x)}{(1-x)^2},\ldots
\end{eqnarray}
and so the improved series for $l_{R\infty}$ is of the form
\begin{equation}
l_{R\infty}=1-x+\sum_{n=1}^\infty \left(\frac{\alpha}{\pi}\right)^n (L_n(x)+
a_{n0}),\qquad x=\frac{\alpha}{\pi}\beta_1 \ln\frac{e\eta}{\gamma \pi m^2}.
\end{equation}
In this series, $L_1$ and $a_{10}=-2\pi^{-2}\zeta'(2)$ are completely
determined, and to determine $L_2$ we need only know the constant $a_{20}$,
since $\beta_3$ is known from the study of the polarization function $\pi_R$;
cf. (48).

   We call attention to the singularities of the functions $L_n(x)$ at the
point $x=1$, which prevent us from using the series (52) near this point, at
which the zeroth approximation for $l_{R\infty}$ vanishes. Unlike the initial
perturbation-theory series, which is applicable for $x\ll 1$, the improved
series (52) is applicable in a wider range of $x$, including values of $x$
that are of the order of $1$ but not too close to unity, viz., for
\begin{equation}
\frac{\alpha}{\pi},\qquad \frac{\alpha}{\pi}\frac{\ln(1-x)^{-1}}{1-x}\ll 1.
\end{equation}
The larger of the quantities on the left is the applicability parameter of
the improved perturbation theory with respect to the radiation field in
the electrodynamics of an intense field with no particles.

\section{ Formation region, pair creation by electric
field and by a particle in a field and other remarks}

   The expressions (1) and (13) for Lagrange function in one- and two-loop
approximations describe the vacuum polarization by constant homogeneous
electromagnetic field.

   The slowly varying field $F$ may be considered as a constant and homogeneous
one if the length $\lambda $ and time $\tau $ on and at which it changes
appreciably is large in comparison with the length and time on and at which the
kinetic energy and momentum of a charge are changed by this field on the
quantities  $mc^2 $ and $mc $:
\begin{equation}
\lambda \gg \frac{mc^2}{eF},\quad \tau \gg \frac{mc}{eF}.
\end{equation}
The length $mc^2/eF$ and time $mc/eF $ may be considered as the length and
time of formation of the process.

Condition (54) coincides with the multiphotonness condition for the processes
in plane wave fields for which the
invariant parameter $\xi=eF\lambda/mc^2 \gg 1$ [35].

  However, for the processes of pair creation by a weak electric field,
when the probabilities are exponentially small, the variation of the field
may be neglected at more hard condition
\begin{equation}
\lambda \gg \frac{mc^2}{e\epsilon}\sqrt{\frac{m^2c^3}{\hbar e\epsilon}}, \qquad
\beta= \frac{\hbar e\epsilon}{m^2c^3} \ll 1,
\end{equation}
see [36].

   In very strong fields the processes are formed on the lengthes of the order
of $\sqrt{\hbar c/eF}$, which are small in comparison with Compton length
$\hbar/mc$, so that the variation of the field may be neglected if
\begin{equation}
\lambda \gg \sqrt{\frac{\hbar c}{eF}}, \qquad \frac{\hbar eF}{m^2c^3} \gg 1,
\end{equation}
see [37],[19].

   These conditions are confirmed in consideration of one-loop Lagrange
function of nonstationary electromagnetic field [38].

   The processes caused in a constant field by a real particle with momentum
$p_{\alpha}$ are defined not only by the field strengthes $\epsilon $ and
$\eta $ but also by the important parameter
\begin{equation}
\chi=\frac{\sqrt{(eF_{\alpha \beta}p^{\beta})^2}}{m^3},
\end{equation}
which becomes the main one for large particle energies $p^0 \gg m$, when
\begin{equation}
\chi \gg \frac{e\epsilon}{m^2},\; \frac{e\eta}{m^2}
\end{equation}
and in a system where $p^0 \sim m $ the field looks as a crossed field.

   The probability of pair creation by a charged particle moving in a crossed
field with momentum $p_{\alpha}$ is proportional to
\begin{equation}
\exp\left(-\frac{16}{3\chi}\right),\quad \hbox{if} \quad \chi \ll 1.
\end{equation}

   For the pair creation of light particles with masses $m'$ by heavy charged
particle with mass $m$ and momentum $p_{\alpha}$ in a crossed field the
probability is proportional to
\begin{equation}
\exp\left(-\frac{2\sqrt3}{\zeta}\right),\quad \hbox{if} \quad \zeta=
\chi \frac{m^2}{m'^2} \ll 1,\quad m'\ll m.
\end{equation}
   The probability of pair creation by a photon with momentum $k_{\alpha}$
in a linearly- or circularly-polarized monochromatic wave with amplitude $F$
and frequency $\omega$ is proportional to
\begin{equation}
\exp\left[-\frac{8}{3\chi}\left(1-\frac{1}{10\xi^2}\right)\right] \quad
\hbox{or} \quad \exp\left[-\frac{8}{3\chi}\left(1-\frac{1}{15\xi^2}\right)
\right],
\end{equation}
if
$$
\chi=\frac{\sqrt{(eF_{\alpha\beta}k^{\beta})^2}}{m^3} \ll 1,\qquad
\xi=\frac{eF}{m\omega} \gg 1,\qquad \xi^2\chi \ge 1.
$$
All exponentials in (59-61) have different exponents and essentially differ
from exponential $\exp(-\pi m^2/e\epsilon)$ defining the probability of pair
production by electric field. See [39,35,36] about these processes.

   Recently two-loop Lagrange functions of a constant field in spinor and
scalar QED were recalculated by Fliegner, Reuter, Schmidt and Schubert [40]
using dimentonal regularization instead of proper-time one. Unfortunately,
their expressions contain only magnetic field, so that very interesting
effects caused by electric field turn out to be omitted.

   The problems connected with the Lagrange function of intensive
electromagnetic field in quantum electrodynamics once more were considered and
analysed by author in a detailed survey [41].

   This work was supported by Russian Fund for Fundamental Research, grant
No 96-02-17314.

\section*{ Appendix}

   Here the functions $K$ and $K_0$ defining the proper-time representation
of ${\cal L}_R^{(2)}$ in spinor electrodynamics are presented:
$$
K(s,s')=e^{-im^2(s+s')}\left\{\frac{e^4\eta^2\epsilon^2}{PP'}\left[4m^2(SS'+
PP')I_0-iI \right]-  \right.  $$
$$
\left.\frac{1}{ss'(s+s')}\left[4m^2-\frac{2i}{s+s'}+\frac{e^2(\eta^2-
\epsilon^2)}{3}\left(2m^2(ss'-2s^2-2s'^2)-\frac{5iss'}{s+s'}\right)\right]
\right\},
$$
$$
K_0(s)=e^{-im^2s}\left(4m^2+i\frac{\partial}{\partial s}\right)\left(
\frac{e^2\eta\epsilon}{\tan e\eta s\:\tanh e\epsilon s}-\frac{1}{s^2}+
\frac{e^2(\eta^2-\epsilon^2)}{3}\right),
$$
$$
I_0=\frac{1}{b-a}\ln\frac ba,\qquad I=\frac{q-p}{b-a}I_0-\frac{qb^{-1}-
pa^{-1}}{b-a},
$$
$$
a=e\eta(\cot e\eta s+\cot e\eta s'),\qquad b=e\epsilon(\coth e\epsilon s+
\coth e\epsilon s'),
$$
$$
p=\frac{2e^2\eta^2\cosh e\epsilon(s'-s)}{\sin e\eta s\:\sin e\eta s'},\qquad
q=\frac{2e^2\epsilon^2\cos e\eta(s'-s)}{\sinh e\epsilon s\:\sinh e\epsilon s'},
$$
$$
P=\sin e\eta s\:\sinh e\epsilon s,\qquad S=\cos e\eta s\:\cosh e\epsilon s.
$$

   The functions $K$ and $K_0$ for ${\cal L}_R^{(2)}$ in scalar electrodynamics
are different but have similar structure, see [20].

\end{document}